# Intense and Tunable Multi-color Terahertz Radiation from Laser-Shaped Electron Beams

Yin Kang [1], Weiyi Yin [1], Xianzhe Li [1,2], Yixuan Liu [1], Yue Wang [3], Yuan Ma [2,4], Kaiqing Zhang [1,2]*, and Chao Feng [1,2]*

[1] *Shanghai Advanced Research Institute, Chinese Academy of Sciences, Shanghai 201210, China*
[2] *University of Chinese Academy of Sciences, Beijing 100049, China*
[3] *Zhangjiang Laboratory, Shanghai 201210, China*
[4] *Shanghai Institute of Applied Physics, Chinese Academy of Sciences, Shanghai 201800, China*

**Abstract** High-power multi-color terahertz (THz) radiation exhibits extraordinary scientific application prospects at various scientific frontiers, for its capacity to deliver THz excitation at multiple frequencies simultaneously. However, the generation of high-power multi-color THz radiation with tunable frequencies remains a challenge for existing techniques. Here, a technique by combining the multi-laser pulses frequency beating and coherent undulator amplification is proposed for generating high-power multi-color THz radiation with tunable frequency. Numerical simulations indicate that the proposed technique can produce multi-color THz radiation with three to six distinguished colors and a peak power up to hundreds of MW, and the temporally separated two-color pulses can also be produced by employing undulators with different resonance. Due to the intrinsic properties of the proposed technique, the THz frequencies, the color number and the frequency interval can be effectively controlled by simply adjusting the beating laser. This method paves the way for advanced application of THz pump-THz probe experiments for selective excitation of atomic multi-level systems and molecular fingerprint recognition.

High-power terahertz (THz) radiation has become a unique tool for the scientific frontier research on low-energy excitations (such as lattice vibrations, spin waves, and internal excitations of bound electron-hole pairs [1-6]), since its photon energy matches well with various low-lying excitation states. With the development of high-power THz source, the generation of multi-color THz radiation (i.e. coherent radiation with several discrete frequency components) has attracted widespread attention, for its extraordinary application prospect in many scientific frontiers, such as coherent THz manipulation of atomic/molecular multi-level systems [7-11], THz tomography [12-15], differential THz absorption lidar [16,17], and THz molecular fingerprint recognition [18-20]. In recent years, the newly developed THz pump-THz probe (TPTP), based on the temporally separated two-color pulses, makes it possible to explore ultrafast dynamical processes in materials science, including insulator-to-metal phase transitions [21], semiconductor carrier dynamics [22-24], effective mass anisotropy of hot electrons [25], ultrafast dynamics in ferroelectrics and coherent electronic excitations in condensed matter systems [26-28]. By combining with THz scanning tunneling microscopy (THz-STM), the TPTP can provide unprecedented access to ultrafast dynamics with atomic resolution in single molecules and nanostructures [29].

The main problem holding back the scientific application of multi-color THz radiation is the lack of synchronized multi-color THz sources. Until now, the multi-color THz are mainly produced by quantum cascade laser (QCL) [30-32], photoconductive antennas [33], optical parametric oscillators [34-36], and difference frequency generation (DFG) based on femtosecond (fs) lasers [37]. However, all of them are struggling with relatively lower peak power, frequency tunability and the control of spectral structure.

Free-electron laser (FEL), based on the undulator radiation, has been a credible technique to generate high-power, narrowband radiation with continuous frequency tunability from THz to X-ray [38-40]. In the X-ray regime, the high-power multi-color FEL has been successively demonstrated by adjusting the undulator resonance and beam manipulation in FEL facilities [41-45]. The development of multi-color X-ray FEL has promoted amounts of advanced scientific experiments, such as resonant inelastic X-ray scattering (RIXS) [46-49], X-ray pump-probe (XPP) experiments [50-52], and high-precision spectroscopic analysis [53-55]. In the THz regime, multi-color THz generation based on a low gain FEL has been proposed, which exhibits a relatively lower power, limited spectral lines and poor frequency tunability [56-58]. Until now, the efficient generation of multi-color THz radiation remains a significant challenge for the existing techniques. In our recent work [59], a THz FEL with continuous band coverage via electron beam tailoring has been demonstrated, which makes it possible to precise control of THz radiation.

In this letter, we propose a high-power multi-color THz generation method by combining the multi-laser pulses frequency beating and coherent undulator amplification techniques. Figure 1 illustrates the layout of the proposed method. An electron beam with a beam length (RMS) of 1.67 picosecond (ps) is first generated by an electron gun, and then the beam is accelerated to 120 MeV by a linac. Subsequently, the beam is injected into a modulator, where the beam interacts with a multi-pulse frequency beating seed laser. Three-dimensional (3D) simulations were performed using the FALCON code [60] and GENESIS [61] based on the typical parameters listed in Table 1.

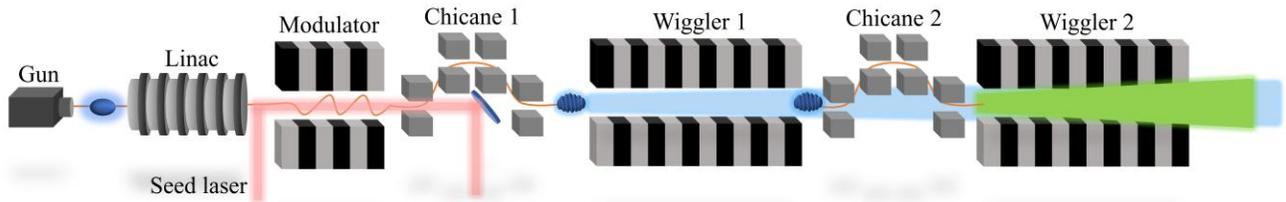

Figure 1. The layout of the proposed method.

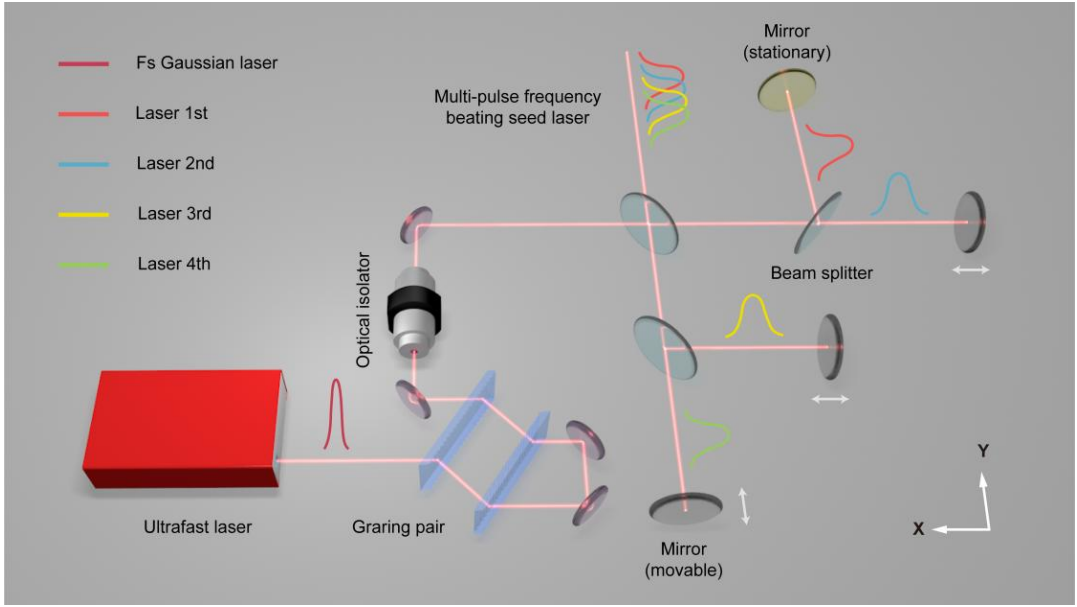

Figure 2. The layout of the multi-pulse frequency beating seed laser system.

Table 1. The main parameters of simulations.

| Parameters | Value |
|---|---|
| Laser wavelength | 800 nm |
| Laser peak power | 48 MW |
| Laser initial pulse duration (FWHM) | 35 fs |
| Grating line | 350 mm$^{-1}$ |
| Grating pair distance | 40 mm |
| Incident angle | 38° |
| Electron beam energy | 120 MeV |
| Energy spread (slice) | 0.01% |
| Beam length (RMS) | 1.67 ps |
| Bunch charge | 4 nC |
| Peak current | 500 A |
| Modulator period | 5 cm |
| Undulator period | 20 cm |

The optical layout of the seed laser system is illustrated in Figure 2. A laser with initial pulse duration $\sigma_0$ (FWHM) of 35 fs and a center wavelength $\lambda_0$ of 800 nm is generated by an ultrafast laser. The electric field of the laser can be expressed as

$$E_{in}(t) = A_0 exp(-\frac{t^2}{\sigma_0^2} + i\omega_0 t), \quad (1)$$

where $A_0$ is the amplitude, $\omega_0$ is the center angular frequency, and $t$ is the longitudinal time coordinate. After a grating pair, the laser pulse is broadened to 3.92 ps with an electric field described by

$$E_{out}(t) = A_0(\frac{\sigma_0}{\sigma_{out}})^{\frac{1}{2}}exp(-\frac{t^2}{\sigma_{out}^2})exp[i(\omega_0 t + \alpha t^2)], \qquad (2)$$

where $\sigma_{out}$ is the broadened pulse duration, and $\alpha = 1/\sigma_0\sigma_{out}$ is the linear chirp parameter. After passing through an optical isolator, the broadened laser pulse is separated as $m$ branches by optical splitters.

Taking four ($m = 4$) branches as an example, the laser branches are delayed 1.33 ps, 2.76 ps, and 4.01 ps separately with respect to the first branches. And then, the four laser branches are recombined by an optical mixer. Finally, a multi-pulse frequency beating seed laser, whose envelope contains multiple discrete THz frequency components (totally $m(m-1)/2$), can be obtained. The beat frequencies can be expressed as

$$f_n = \frac{\alpha\tau_{i,j}}{\pi}, \qquad (3)$$

where $\tau_{i,j}$ is the time delay between the two branches $i$ and $j$. By adjusting the relative time delay of the laser branches, the beating frequencies can be tuned continuously and independently. Considering the intensity of the frequency components, $m-1$ envelope frequencies produced by the two adjacent pulses can be amplified in the following undulator. Figure 3a presents the intensity profiles of the four laser branches, Figure 3b and Figure 3c show the intensity profiles of the multi-pulse frequency beating seed laser and its Fourier transform spectrum.

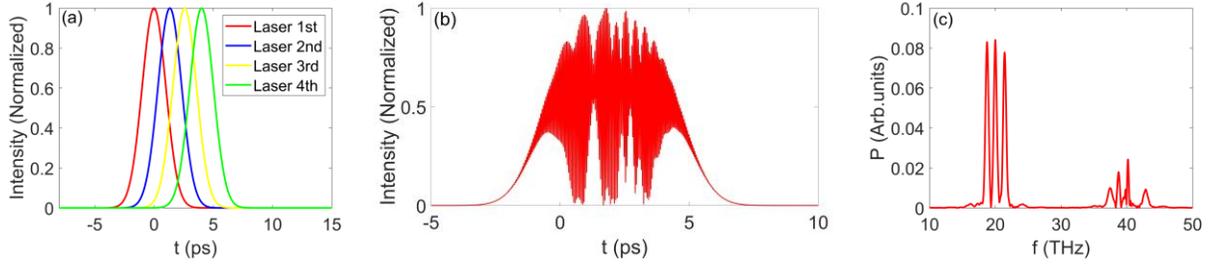

Figure 3. The intensity profiles of the four laser branches (a), the intensity profiles of the multi-pulse frequency beating seed laser (b) and its envelope spectrum (c).

According to Figure 3, the envelope frequencies formed by two adjacent pulses (lasers 1 and 2 (18.75 THz (16 μm)), lasers 2 and 3 (20 THz (15 μm)), lasers 3 and 4 (21.42 THz (14 μm)) dominate the spectrum, while other envelope frequencies have much lower intensity due to large time delays. The gain bandwidth (RMS) of the undulator can be expressed as [62-64]

$$\sigma_{\Delta\lambda/\lambda} = \frac{\sqrt{3\sqrt{3}\rho}}{k_u z}, \quad (4)$$

where $\lambda$ is the resonant wavelength, $\rho$ is the Pierce parameter, $k_u$ is the undulator wave vector, and $z$ is the longitudinal position coordinate along the undulator. For a typical laser beam interaction, the seed laser with a wavelength within $6\lambda\sigma_{\Delta\lambda/\lambda}$ can interact with the electron beam in the modulator. For the proposed technique, the wavelengths of the four laser branches are still 800 nm, which means that the multi-color THz modulation is not constrained by the gain bandwidth of the modulator.

Here, a planar undulator with a dimensionless parameter $K$ of 1.24 and a magnetic period $\lambda_u$ of 5 cm is adopted as the modulator. The dimensionless energy deviation of the beam after the modulator can be linearly expressed as

$$p = p_0 + A\{cos(\omega_0 t + \alpha t^2) + cos[\omega_0(t + \tau_{12}) + \alpha(t + \tau_{12})^2] + cos[\omega_0(t + \tau_{13}) + \alpha(t + \tau_{13})^2] + cos[\omega_0(t + \tau_{14}) + \alpha(t + \tau_{14})^2]\}, \quad (5)$$

where $p_0$ is the initial dimensionless energy deviation, and $A$ is the modulation amplitude. Then, the beam passes through a dispersion section (chicane 1) with a momentum compression factor $R_{56} = 7.16$ mm to convert the energy modulation into density modulation, and the longitudinal position $s$ of the electron becomes

$$s = s_0 + R_{56} p \frac{\sigma_\gamma}{c\gamma_0}, \quad (6)$$

where $s_0$ is the initial longitudinal position along the beam, $\sigma_\gamma$ is the energy spread, $c$ is the speed of light in vacuum, and $\gamma_0$ is the average energy of the beam. After the dispersion section, the longitudinal phase space, the current profile, and the bunching factor distribution are shown in Figure 4. One can find that the beam contains microbunches with multiple distinct THz frequencies (18.72 THz, 20.02 THz, and 21.41 THz) along the longitudinal direction with a maximum bunching factor of 0.32.

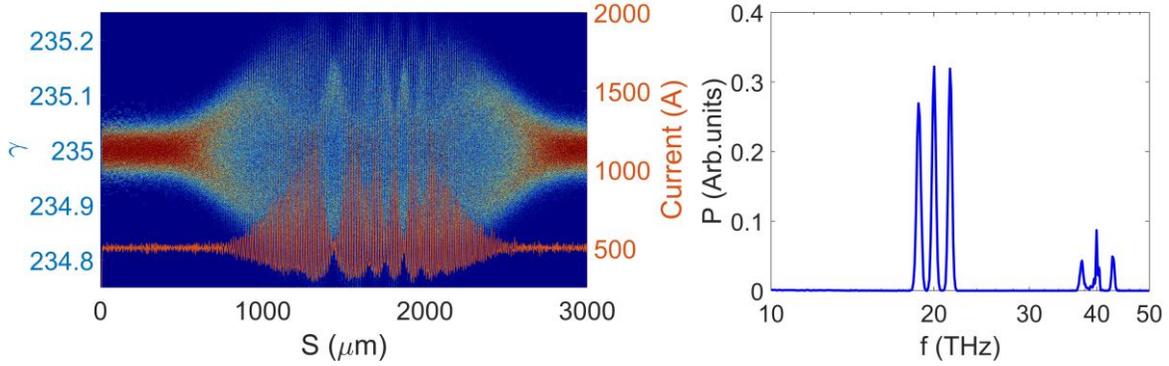

Figure 4. The longitudinal phase space (left), current profile (left), and the bunching factor distribution (right) of the beam after density modulation.

After that, the modulated beam is sent into a planar undulator with $K$ of 3.81 and $\lambda_u$ of 20 cm to generate multi-color THz radiation. The undulators 1 and 2 firstly resonate at the same frequency (20 THz), and the chicane 2 is turned off. Figure 5 shows the peak power and bunching factor evolutions of multi-color THz radiation. To show the frequency tunability of the proposed technique, the multi-color THz radiation with different frequency intervals is also given in Figure 5. From Figures 5a, 5b and 5c, one can observe that the peak power of the three-color THz radiation reaches 0.58 GW at 7.6 m, with a maximum bunching factor of 0.85. The spectrum has three discrete frequency components (18.74 THz, 20.05 THz, 21.46 THz). The radiation frequencies and frequency intervals can be easily tuned by simply adjusting the time delay of the laser branches. In addition, the simulated results of THz radiation with six colors are also presented in Figures 5d, 5f and 5e. According to these results, the six-color THz radiation is obtained with a peak power of 0.43 GW at 5 m. The spectrum exhibits six distinct frequency components (17.24 THz, 18.04 THz, 18.94 THz, 20.05 THz, 21.16 THz and 22.36 THz), and the frequency intervals can be easily controlled by adjusting the optical delays.

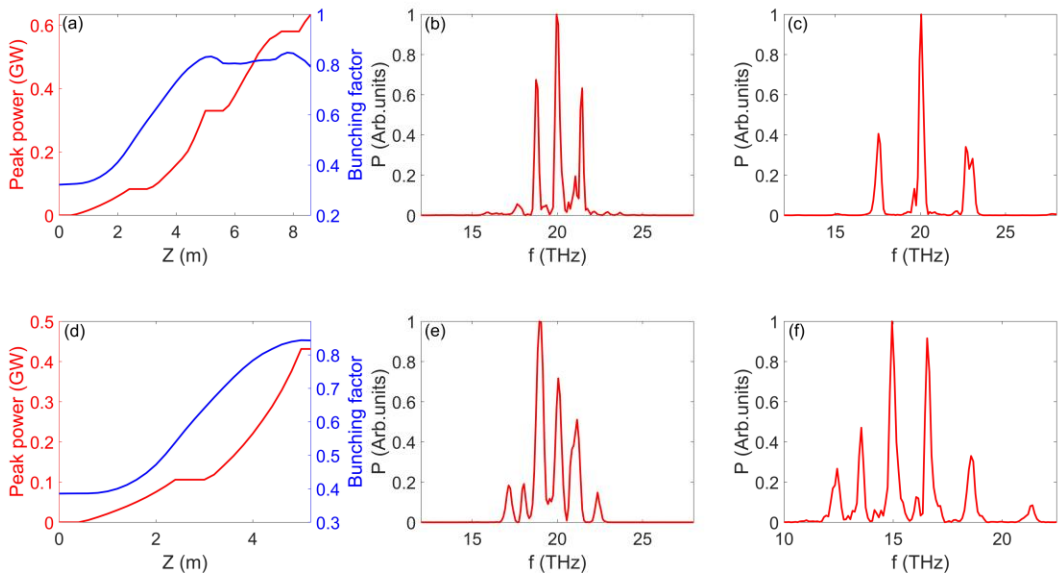

Figure 5. The evolutions of peak power and bunching factor (a) of three-color THz radiation, and the spectra of narrow (b) and wide intervals (c). The evolutions of peak power and bunching factor (d) of six-color THz radiation, and the spectra of narrow (e) and wide intervals (f).

However, the frequency components closed to $6\lambda\sigma_{\Delta\lambda/\lambda}$ are difficult to be amplified, since that the resonance deviates the undulator natural linewidth. The natural linewidth of the undulator can be expressed as

$$\Delta\lambda = \frac{\lambda}{nN}, \quad (7)$$

where $n$ is harmonic number, and $N$ is the period number of the undulator. The minimum interval between discrete THz wavelengths should be greater than $\Delta\lambda$, otherwise longitudinal modes will merge [56]. Ultimately, the longitudinal mode differences $l_{mode}$ of radiation must satisfy

$$\frac{\lambda}{nN} < l_{mode} < 6\lambda\frac{\sqrt{3\sqrt{3}\rho}}{k_u z}. \quad (8)$$

To overcome the limitation of undulator gain bandwidth and satisfy the requirements of the THz pump-THz probe (TPTP) experiments, the temporally separated and independent amplification of two-color THz pulses has also been considered. Here, the different longitudinal positions of the electron beam are modulated with two laser pulses involved two discrete beating frequency. For this case, laser branches 1 and 2 are combined at the head of the electron beam, the beating laser involved the Fourier components of 20 THz (15 μm) interacts with the head of beam in modulator. Simultaneously, laser branches 3 and 4 (beating frequency is 30 THz (10 μm)) are combined at the tail of the electron beam and interact with the tail of beam. Subsequently, the modulated beam passes through the chicane 1 with $R_{56} = 4.77$ mm to convert energy modulation into density modulation. The longitudinal phase space, current profile, and bunching factor distribution of the beam after density modulation are shown in Figure 6. One can find that the beam has a 20 THz microbunching structure with a bunching factor of 0.22 at the head and a 30 THz microbunching structure with a bunching factor of 0.27 at the tail.

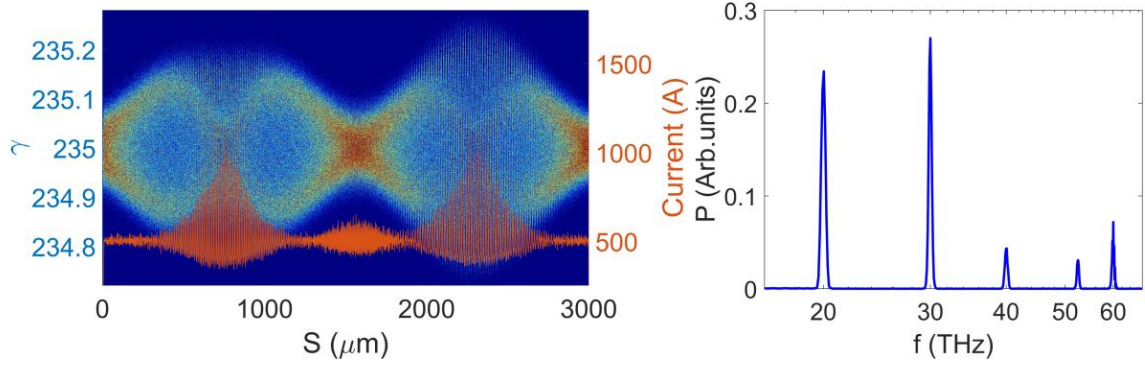

Figure 6. The longitudinal phase space (left), current profile (left), and bunching factor distribution (right) of the beam after density modulation.

Then, the modulated beam is sent into undulator 1 with $K$ of 3 and $\lambda_u$ of 20 cm to generate 30 THz radiation. Figure 7 shows the evolutions of peak power and bunching factor of 30 THz radiation, the spectrum, and longitudinal phase space of the beam after radiation. According to Figure 7, it can be observed that the peak power reaches 0.24 GW at 7.6 m, the bunching factor reaches 0.8, and the spectral bandwidth (FWHM) is 0.11 μm. More importantly, the energy spread at the tail of the beam increases due to radiation, while the 20 THz microbunching structure at the head is preserved.

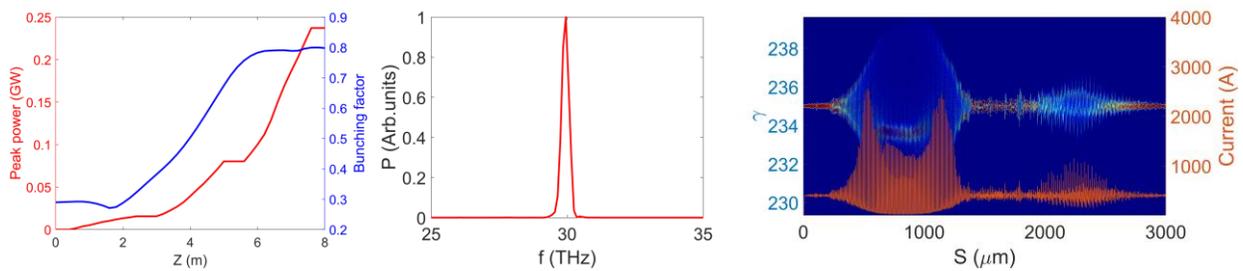

Figure 7. The evolutions of peak power and bunching factor (left) of 30 THz radiation, the spectrum (middle), and longitudinal phase space (right) of the beam after radiation.

In this case, the chicane 2 with $R_{56} = 1.2$ mm is used to accurately control the time interval of temporally separated two-color THz pulses, the time delay can be adjusted between hundreds of fs ($R_{56} = 1.2$ mm) to 3 ps ($R_{56} = 0$ mm). And then, the beam is sent into undulator 2 with $K$ of 3.81 and $\lambda_u$ of 20 cm to generate 20 THz radiation. The evolutions of peak power and bunching factor of 20 THz radiation and the spectrum are shown in Figure 8. Due to the time delay induced by chicane 2 and the slippage effect, the THz radiation generated at the tail of the beam will catch up with the head of the beam, thereby the two THz pulses may have a pulse overlap. Through the coordination of the chicane 1 and 2, the 20 THz structural overbunching of the beam can be avoided. However, due to the overall increase in energy spread of the beam after the 30 THz radiation, the 20 THz bunching factor is slightly reduced from 0.22 to 0.18. Finally, the peak power of 20 THz radiation reaches 0.52 GW at 15.2 m, and the spectral bandwidth is 0.2 μm.

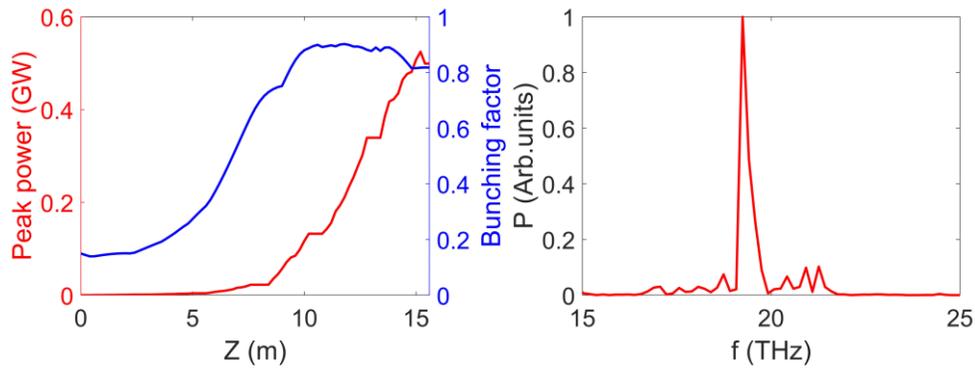

Figure 8. The evolutions of peak power and bunching factor (a) of 20 THz radiation and the spectrum (b).

In conclusion, a novel technique is proposed and demonstrated to generate high-power multi-color THz radiation by combining the multi-laser pulses frequency beating and coherent undulator amplification technique. To satisfy the requirements of the THz pump-THz probe (TPTP) experiments, the temporally separated and independently amplified two-color THz pulses are also considered. For the proposed technique, the THz frequencies, the color number and the frequency interval can be effectively controlled by simply adjusting the beating laser. The 3D numerical simulations are carried out, and the results show that the proposed technique can generate discrete narrowband THz radiation with three to six colors and a peak power of hundreds of MW level.

The proposed technique can also be used to generate temporally separated two-color THz pulses (20 THz, 30 THz) with peak powers at hundreds of MW level and a tunable delay from hundreds of fs to several ps by modulating the beam at different longitudinal positions and amplifying the two pulses separately. The THz radiation source proposed here can promote diverse scientific applications in fundamental and applied physics fields such as atomic multi-level excitation, molecular fingerprint recognition, TPTP, from THz spectroscopy to ultrafast dynamic processes.

## Acknowledgement

The authors would like to thank Hong Qi for useful discussions on FEL physics and simulations. This work is supported by the National Natural Science Foundation of China (12275340, 12435011, 12105347), CAS Project for Young Scientists in Basic Research (YSBR-115), Shanghai Municipal science and Technology Major Project and Innovation Program of Shanghai Advanced Research Institute, CAS (2024CP001).